**Unconventional Scaling of the Anomalous Hall Effect Accompanying Electron**

**Localization Correction in the Dirty Regime**

Y. M. Lu,[1] J. W. Cai,[1,*] Zaibing Guo,[2] and X. X. Zhang[2]

[1]*Beijing National Laboratory for Condensed Matter Physics, Institute of Physics,*

*Chinese Academy of Sciences, Beijing 100190, People's Republic of China*

[2]*Core Labs, King Abdullah University of Science and Technology (KAUST),*

*Thuwal 23955-6900, Saudi Arabia*

Scaling of the anomalous Hall conductivity to longitudinal conductivity, $\sigma_{AH} \propto \sigma_{xx}^2$, has been observed in the dirty regime of two-dimensional weak and strong localization regions in ultrathin, polycrystalline, chemically disordered, ferromagnetic FePt films. The relationship between electron transport and temperature reveals a quantitatively insignificant Coulomb interaction in these films while the temperature dependent anomalous Hall conductivity experiences quantum correction from electron localization. At the onset of this correction, the low-temperature anomalous Hall resistivity begins to be saturated when the thickness of the FePt film is reduced, and the corresponding Hall conductivity scaling exponent becomes 2, which is above the recent unified theory of 1.6 ($\sigma_{AH} \propto \sigma_{xx}^{1.6}$). Our results strongly suggest that the correction of the electron localization modulates the scaling exponent of the anomalous Hall effect.

The anomalous Hall effect (AHE) in ferromagnetic conductors, though first discovered by E. H. Hall in 1881, has received renewed interest in recent years due to its close connection with spin transport phenomena [1] and the controversial mechanisms [2]. It is now generally accepted that there are three mechanisms responsible for AHE. The first two involve the extrinsic mechanisms, namely skew scattering and side jump, both of which result from the asymmetric impurity scattering caused by the spin-orbit interaction and yield scaling relations between the anomalous Hall resistivity, $\rho_{AH}$ (Hall conductivity $\sigma_{AH}$), and longitudinal resistivity, $\rho_{xx}$ (conductivity $\sigma_{xx}$), as $\rho_{AH} \propto \rho_{xx}$ ($\sigma_{AH} \propto \sigma_{xx}$) and $\rho_{AH} \propto \rho_{xx}^2$ ($\sigma_{AH} \sim const.$), respectively [3,4]. The other is the intrinsic mechanism, which arises from the transverse velocity of the Bloch electrons induced by the spin-orbit interaction together with interband mixing, and also gives $\rho_{AH} \propto \rho_{xx}^2$ ($\sigma_{AH} \sim const.$) [5]. Nowadays, this intrinsic AHE has been reinterpreted in terms of the Berry curvature of the occupied Bloch states [2, 6].

Recently, a unified theory of AHE for multiband ferromagnetic metals with diluted impurities has been proposed and later verified by experimental results from a variety of itinerant ferromagnets [7-12]. It predicts three distinct scaling regimes in the AHE that are functions of conductivity. In the clean regime ($\sigma_{xx} > 10^6$ S/cm), the skew scattering mechanism dominates, resulting in $\sigma_{AH} \propto \sigma_{xx}$. With decreasing conductivity to the intermediate disorder regime, ($\sigma_{xx} \sim 10^4 - 10^6$ S/cm), the intrinsic contribution becomes dominant, yielding $\sigma_{AH} \sim const.$ In the dirty regime with high disorder ($\sigma_{xx} < 10^4$ S/cm), the intrinsic contribution is strongly decayed, resulting in a scaling relation expressed as $\sigma_{AH} \propto \sigma_{xx}^{\gamma}$ with $\gamma \sim 1.6$, irrespective of the different mechanisms responsible for the electron transport. It should be emphasized that, in the dirty regime, the physics of AHE can be enriched by the quantum effects of the Coulomb interaction (CI) and disorder-induced electron localization (EL) [2, 13-15], which the unified theory does not take into account [7, 8]. On this open issue, previous experimental studies mostly focused on the quantum corrections to $\sigma_{AH}$ through altering the temperature of films with given thicknesses [16-19]. The evolution of the scaling exponent, $\gamma$, particularly in the presence of the quantum correction to $\sigma_{AH}$, remains poorly known [2]. Interestingly, it has been recently reported that the low-temperature AHE scaling in CNi$_3$ films exhibits a peculiar $\sigma_{AH} \propto \sigma_{xx}^2$ relationship near the Mott-Anderson critical region, and CI was suggested to play a crucial role in the unconventional scaling exponent [20]. In this Letter, we seek to elucidate how EL modulates the scaling exponent of $\sigma_{AH} \propto \sigma_{xx}^{\gamma}$ in homogenous polycrystalline FePt films, which exhibit ideal features of negligible CI and well-defined quantum correction to $\sigma_{AH}$ from EL. Our findings clarify the role of EL in affecting the AHE scaling and point to an experimental approach that can verify the role that EL plays.

We deposited SiO$_2$(5 nm)/FePt(1.2-100 nm)/SiO$_2$(10 nm) films on thermally oxidized Si substrates at room temperature by RF/DC magnetron sputtering. The equiatomic FePt alloy layers formed a homogenously continuous structure, which was confirmed by transmission electron microscopy observations [21]. The x-ray diffraction (XRD) pattern revealed the face-centered cubic polycrystalline structure of the FePt films (chemically disordered). Transport measurements were carried out on a Quantum Design PPMS-14H at 300-2 K. In studying Hall effect, the offset signal was removed, and $\rho_{AH}$ was obtained as the zero field extrapolation of the high field $\rho_{xy}(H)$. The magnetization of all samples measured by MPMS-XL yielded the bulk value of FePt and changes very little in the low temperature range. A detailed description of the film's growth, structural characterization, and sample preparation for the transport measurements can be found in our previous paper [21].

The inset in Fig. 1(a) shows the dependence of thickness on sheet resistance, $R_{xx}$, measured at 300 K. The $R_{xx}$ increased by about five orders of magnitude from 8.57 $\Omega$ to 17.6 k$\Omega$, when the film thickness ($t_{FePt}$) was reduced from 100 nm to 1.2 nm. The corresponding $\rho_{xx}$ also monotonically increased and fell in the range of 86-2100 $\mu\Omega$ cm. The normalized sheet resistance, $R_{xx}(T)/R_{xx}(300\,\mathrm{K})$, as a function of temperature, $T$, is presented in Fig. 1(a). In relation to this figure, we highlight that the FePt films that were 20 nm or thicker exhibited typical metallic behaviors ($\mathrm{d}R_{xx}/\mathrm{d}T > 0$) across the entire temperature region (300-2 K) and their normalized sheet resistance curves were identical. Fig. 1(a) shows that the resistance of the 20 nm FePt film tends to decline very slowly at temperatures below 30 K with the residual resistance ratio, $R_{xx}(2\,\mathrm{K})/R_{xx}(300\,\mathrm{K})$, as large as ~0.73, which might be ascribed to the intensive atomic disorder scattering in chemically disordered FePt films. In fact, the residual $\rho_{xx}$ value of these bulk FePt films ($t_{FePt} \geq 20\,\mathrm{nm}$) reached ~60 $\mu\Omega$ cm, significantly exceeding that of pure iron (~5 $\mu\Omega$ cm) or $L1_0$-ordered FePt (~15 $\mu\Omega$ cm) films [22, 23]. When the FePt thickness was thinner than 12 nm, the $R_{xx}$ indicated insulating characteristics ( $dR_{xx}/dT < 0$ ) in the low-temperature region. An appreciable upturn in $R_{xx} \sim T$ is evident in very thin films ($t_{FePt} \leq 3$ nm). Quantitative analysis of the insulating resistance results shows that a linear logarithmic temperature dependence of $R_{xx}$ emerges at $1.6$ nm $\leq t_{FePt} \leq 3$ nm. When $t_{FePt}$ was further decreased to 1.4-1.2 nm, the $R_{xx}$ varied faster than the $\ln T$ law in the low-temperature region. Actually it was appropriate to fit the $R_{xx}$ of the 1.2 nm FePt film with a variable range hopping type of conduction [24], which will be detailed later.

As mentioned above, when the FePt thickness is reduced to 3-1.6 nm, $R_{xx}$ changes linearly with $\ln T$ at low temperatures. This phenomenon can be ascribed to the two-dimensional (2D) weak EL and/or the 2D CI [13]. To clarify if CI has a quantum correction to $R_{xx}$ or sheet conductance, $G_{xx}$ ( $1/R_{xx}$ ), in the weakly localized region, we analyzed the low-temperature conductivity quantitatively. In the 2D case, $G_{xx}$ can be expressed as

$$G_{xx}(T) = G_0 + p\frac{e^2}{\pi h}\ln\left(\frac{T}{T'}\right) + \left(1-F\right)\frac{e^2}{\pi h}\ln\left(\frac{T}{T'}\right), \tag{1}$$

where the first term on the right side of the equation stands for the Drude conductance, the second term is the contribution of weak EL, and the third term is the CI correction [13, 14]. In Eq (1), the parameter $F$ is a measure of the screening with $0 \leq F \leq 1$, and $p$ is the temperature exponent of the inelastic scattering length, $l_{in} \sim T^{-p/2}$. The value of $p$ is governed by inelastic relaxation mechanisms: $p = 1$ for the electron-electron scattering, whereas $p = 2$, 3, or 4, depending on the material and the temperature, for the electron-phonon scattering ([13] and references therein). Figure 1(b) shows the $G_{xx}$ obtained in a 1.8 nm FePt film as a function of $\ln T$. The perfect linear behavior for the data in the 2-20 K range on a logarithmic scale verifies the validity of the 2D assumption [13, 14]. The fitting to the linear part of the $G_{xx} \sim \ln T$ curve yields a slope of $1.2 \times 10^{-5}$ S, i.e., the value of $e^2/\pi h$. This leads to $p + (1 - F) = 1$, namely either $p = F = 0$ or $p = F = 1$. Considering that the AHE conductivity of the thin films receives a quantum correction from the weak EL, which will be shown later, the case of $p = 0$ and $1 - F = 1$ can be ruled out. We thus have $p = 1$ and $1 - F = 0$. This suggests that the CI correction to the Drude conductance is negligible. We have also fitted the data of other films with 1.6 nm$\leq t_{FePt} \leq$3 nm and found that all the fitted slopes of $G_{xx}$ vs $\ln T$ converge nicely to the common value $e^2/\pi h$. Therefore, the absence of the CI correction to the Drude conductance is universal in weakly localized FePt thin films.

In the 2D strongly localized case, electrons hop between localized states and the $G_{xx}$ is given by

$$G_{xx}(T) = C \exp[-\left(T^{"}/T\right)^{\beta}], \tag{2}$$

where the exponent $\beta$ depends whether the hopping process is influenced by CI [24-26]. $\beta = 1/3$ corresponds to Mott's variable range hopping law whereas $\beta = 1/2$ indicates the formation of a Coulomb gap resulting from strong electron correlation [26]. Figure 1(c) displays the $\ln G_{xx}$ of the 1.2 nm FePt film as a function of $T^{-1/3}$ along with the fitting line using Eq (2) with $\beta = 1/3$ at 2-20 K ($C = 0.065$ mS, $T^{"} = 0.372$ K). The excellent fitting result corroborates the assumption that $\beta = 1/3$ in Eq. (2). It is worth noting that Mott's variable range hopping can exist up to 150 K, which might be due to the phonon assisted hopping [27]. Even so, we have still tried to fit the data of 2-20 K using $\beta = 1/2$. The difference between the experimental points and the fitting result is not significant at 2-20 K, but the discrepancy becomes obvious above 50 K. It seems difficult to distinguish the minor difference in power laws applying $\beta = 1/3$ and $\beta = 1/2$ over the limited temperature range of 2-20 K, which suggests that CI cannot be totally ruled out in the strongly localized 1.2 nm FePt film. Anyway, CI correction to $G_{xx}$ can be considered to be insignificant at least in the weakly localized region in FePt films. Parenthetically, the perfect-fit exponent, $\beta$, is 1/5 and 1/4 for 1.4 and 1.3 nm FePt films, respectively, reflecting the crossover from the weak EL limit to the strong EL in the metal-insulator transition [28].

We now turn to the EL correction to anomalous Hall conductance, $G_{\mathrm{AH}}$, by changing temperature of the individual thin FePt films. In the weakly localized region, anomalous Hall resistance, $R_{\mathrm{AH}}$, at low temperatures (2-20 K) varies linearly with $\ln T$, similar to its $R_{xx}$ or $G_{xx}$ vs $T$ behavior. Following Bergmann and Ye's notation [16], normalized relative changes, $\Delta^{\mathrm{N}} R_{xx}$, $\Delta^{\mathrm{N}} R_{\mathrm{AH}}$, $\Delta^{\mathrm{N}} G_{\mathrm{AH}}$, have been used to represent weak EL corrections to $R_{xx}$, $R_{\mathrm{AH}}$ and $G_{\mathrm{AH}}$ ($G_{\mathrm{AH}} = R_{\mathrm{AH}} / (R_{xx}^2 + R_{\mathrm{AH}}^2)$), respectively, and are expressed as

$$\Delta^{\mathrm{N}} R_{xx} = \left( \frac{\pi h}{e^2} \frac{1}{R_0} \right) \left( \frac{\delta R_{xx}}{R_{xx}} \right) = -A_{\mathrm{R}} \ln\left( \frac{T}{T_0} \right), \qquad (3)$$

$$\Delta^{\mathrm{N}} R_{\mathrm{AH}} = \left( \frac{\pi h}{e^2} \frac{1}{R_0} \right) \left( \frac{\delta R_{\mathrm{AH}}}{R_{\mathrm{AH}}} \right) = -A_{\mathrm{AH}} \ln\left( \frac{T}{T_0} \right), \qquad (4)$$

for $|\delta R_{xx}| \ll R_0$ and $R_{\mathrm{AH}}(T) \ll R_{xx}(T)$, which are true for all our films. $\Delta^{\mathrm{N}} G_{\mathrm{AH}}$

could be deduced as

$$\Delta^{\mathrm{N}} G_{\mathrm{AH}} = \left(\frac{\pi\hbar}{e^2}\frac{1}{R_0}\right)\left(\frac{\delta G_{\mathrm{AH}}}{G_{\mathrm{AH}}}\right) = \left(2A_{\mathrm{R}} - A_{\mathrm{AH}}\right)\ln\left(\frac{T}{T_0}\right),\qquad(5)$$

where $R_0$ is the sheet resistance at reference temperature $T_0$ [18, 19]. Using $T_0 = 2\,\mathrm{K}$ and $R_0 = R_{xx}(2\,\mathrm{K})$, for a 1.8 nm FePt film, the normalized relative changes could fit well linearly with $\ln T$ below 20 K, as shown in Fig. 2(a). The fitting gives the prefactors as $A_{\mathrm{R}} = 0.944$, $A_{\mathrm{AH}} = 1.108$ and $2A_{\mathrm{R}} - A_{\mathrm{AH}} = 0.781$. In Fig. 2(b), we summarize the prefactors for weakly localized films (1.6 nm$\leq t_{\mathrm{FePt}} \leq$3 nm). Note that $A_{\mathrm{R}}$ is $0.92 \pm 0.04$ (~1), in excellent agreement with previous results in 2D weakly localized films with homogenous structures [16, 18, 29]. Most interesting, the prefactor of $\Delta^{\mathrm{N}} G_{\mathrm{AH}}$, namely $2A_{\mathrm{R}} - A_{\mathrm{AH}}$, is far from zero in the weakly localized region, indicating that the weak EL correction to $G_{\mathrm{AH}}$ is nonzero. We also note that the value of $2A_{\mathrm{R}} - A_{\mathrm{AH}}$ increases with increasing disorder strength. Due to the fact that CI correction to $G_{\mathrm{AH}}$ vanishes for both the skew scattering and the side jump and that weak EL correction to $G_{\mathrm{AH}}$ is zero for side jump but nonzero for skew scattering [29, 30, 31], the monotonic increase in the $2A_{\mathrm{R}} - A_{\mathrm{AH}}$ with reducing $t_{\mathrm{FePt}}$ from 3 to 1.6 nm can be attributed to the increased weight of the skew scattering in the AHE as observed in homogenous Fe films [18]. It is worth noting that for the weakly localized 12-6 nm FePt films, $\Delta^{\mathrm{N}} G_{\mathrm{AH}} \approx 0$, namely the quantum correction to $G_{\mathrm{AH}}$ is negligible.

As for the strongly localized film ($t_{\mathrm{FePt}} = 1.2\,\mathrm{nm}$), similar to its $G_{xx}$ or $R_{xx}$ vs $T$ behavior [Fig. 1(c)], a hopping form $\ln R_{\mathrm{AH}} \sim T^{-\beta}$ with $\beta = 1/3$ at low temperatures was observed as shown in the inset of Fig. 3. The normalized relative changes, $\Delta^N R_{xx}$, $\Delta^N R_{AH}$ and $\Delta^N G_{AH}$, with respect to their reference values at 2 K are displayed in Fig. 3. Note that $\Delta^N R_{xx}$, $\Delta^N R_{AH}$ or $\Delta^N G_{AH}$ does not have a nice linear relationship with $\ln T$ even below 20 K. Most remarkably, the $\Delta^N G_{AH}$ value quickly deviates from zero as the temperature increases from the reference one (2 K). For the 1.4 and 1.3 nm thick FePt films, nonzero $\Delta^N G_{AH}$ was also observed at temperatures other than 2 K. It is evident that there is a significant quantum correction to $G_{AH}$ in the strong localization region, although it is difficult to evaluate the CI and EL corrections to $G_{AH}$ separately in the hopping conduction regime.

With well-confirmed quantum corrections to the AHE in the individual thin FePt films, the scaling behavior of AHE at low temperatures is further examined by varying the thickness of the film (accordingly altering the disorder strength). To gain a complete view of the thickness and temperature dependences of $\rho_{AH}$ ( $R_{AH} \cdot t$ ), Fig. 4(a) displays the dependence of $\rho_{AH}$ on $\ln T$ for some representative samples. The metallic films with thicknesses between 20 and 100 nm always show a decrease in the $\rho_{AH}$ with decreasing temperature, but the $\rho_{AH}$ in weakly and strongly localized films displays $\rho_{AH} \sim \ln T$ and $\ln(\rho_{AH}) \sim T^{-\beta}$ tendencies at low temperatures (also see Fig. 2(a) and the inset of Fig. 3). Figure 4(b) shows the $\rho_{AH}$ vs $\rho_{xx}$ plot at 5 K and 2 K for films of different thicknesses. $\rho_{AH}$ increases with the rise in $\rho_{xx}$ as $t_{FePt}$ decreases from 100 to 3 nm, mainly caused by the surface scattering, but when $t_{FePt}$ further decreases, $\rho_{AH}$ becomes saturated while $\rho_{xx}$ continues to rise. As shown in Fig. 4(b), the $\rho_{AH}$ maintains an almost constant value of ~ 4.4 μΩ cm in an appreciably broad $\rho_{xx}$ region (110-3000 μΩ cm), corresponding to $1.2\,\text{nm} \leq t_{\text{FePt}} \leq 3\,\text{nm}$ . This disorder-independent $\rho_{\text{AH}}$ introduces a significant deviation from the unified theory. Figure 4(c) shows $\sigma_{\text{AH}}\,(G_{\text{AH}}/t)$ as a function of $\sigma_{xx}$ at 2 K. For films with metallic behaviors ( $20\,\text{nm} \leq t_{\text{FePt}} \leq 100\,\text{nm}$ ) but a high residual resistivity, $\sigma_{xx}$ is about $1.58 \times 10^4\,\text{S/cm}$ and $\sigma_{\text{AH}} \sim 780\,\text{S/cm}$, regardless of the thickness of the film. The value of $\sigma_{\text{AH}}$ is close to the theoretical prediction of the intrinsic contribution of the order of $e^2/ha \sim 10^3\,\text{S/cm}$ (with a lattice constant of $a \approx 4$ Å) in the 3D intermediate disorder regime [7, 8]. We hence argue that the AHE in metallic FePt films is dominated by the resonant intrinsic Berry-phase contribution. For films with thicknesses in the range of 12-6 nm, the magnitude of $\sigma_{xx}$ ( $1.46-1.25 \times 10^4\,\text{S/cm}$ ) seems to drop into the critical regime between intermediate and high disorder, and the $\sigma_{\text{AH}}$ gradually decreases from 700 to 590 S/cm. As $\sigma_{xx}$ passes through this crossover and goes into the dirty regime ( $\sigma_{xx} \leq 0.91 \times 10^4\,\text{S/cm}$ ) at $t_{\text{FePt}} \leq 3\,\text{nm}$ , where $G_{\text{AH}}$ (or $\sigma_{\text{AH}}$ ) is well demonstrated to receive a quantum correction, the $\sigma_{\text{AH}}$ scales as $\sigma_{\text{AH}} \sim \sigma_{xx}^{\gamma}$ with $\gamma = 2$ . This is in striking contrast with the universal scaling exponent, $\gamma \sim 1.6$ . It should be pointed out that the scaling exponent $\gamma = 2$ has been observed in ultrathin CNi$_3$ films with 2D strong EL near the threshold of the metal-insulator transition ( $k_F l \sim 4.00 - 0.74$ ), where the disorder-enhanced CI was taken as the key to modulating the scaling exponent of AHE [20]. In our FePt films, the scaling relation, $\sigma_{\text{AH}} \sim \sigma_{xx}^2$, holds in a wide dirty regime ranging from weak to strong localization ( $k_F l \sim 70.9 - 0.95$ in Ref. [28]). Moreover, the CI is negligible in the weakly localized regime of FePt films as demonstrated in the electron transport characterization, which excludes the possibility of CI modulating the scaling exponent of AHE. Most important, the emergence of an unconventional scaling exponent, $\gamma = 2$, starting at $t_{FePt} = 3\,nm$ coincides with the onset of the quantum correction to AHE by changing the temperature, strongly suggesting that the electron localization correction modulates the AHE scaling exponent at low temperatures.

In conclusion, we have found an unconventional low-temperature AHE scaling ($\sigma_{AH} \propto \sigma_{xx}^2$) in polycrystalline FePt films in the dirty regime, which is accompanied by a quantum correction to $\sigma_{AH}$ in both weak and strong EL regions. Since the electron transport characterization reveals a negligible CI in localized films, the unique scaling relation is very likely due to the emergence of the EL correction to AHE rather than CI. Detailed theoretical studies are needed to understand how AHE is influenced by electron localization.

This work was supported by the National Basic Research Program of China under Grant No.2009CB929201 and the National Natural Science Foundation of China under Grant Nos.51171205, 51021061 and 50831002. Dr. Virgina Unkefer improved the English.

# Reference

[*]Corresponding author. E-mail: jwcai@iphy.ac.cn

**Figure captions:**

FIG. 1. (color online). (a) The normalized sheet resistance as a function of $\ln T$ for FePt films with 20, 12, 6, 3, 2.2, 1.8, 1.6, 1.4, 1.3, and 1.2 nm thicknesses. Inset: Log-log plot of the sheet resistance as a function of the FePt thickness at 300 K. (b) Sheet conductance versus the logarithm of temperature for a 1.8 nm FePt film. (c) Logarithm of sheet conductance as a function of $T^{-1/3}$ for a 1.2 nm-thick FePt film. The straight lines in (b) and (c) are the linear fit to the data between 2 and 20 K.

FIG. 2. (color online). (a) $\ln T$ dependence of normalized relative changes in the sheet resistance, $R_{xx}$, anomalous Hall resistance, $R_{AH}$, and anomalous Hall conductance, $G_{AH}$, for a 1.8 nm FePt film. The straight lines are the linear fit to the data between 2 and 20 K. (b) Prefactors $A_R$, $A_{AH}$, and $2A_R - A_{AH}$ as a function of sheet resistance, $R_{xx}$, of 2 K in the weak localization region.

FIG. 3. (color online). The normalized relative changes as a function of $\ln T$ for a 1.2 nm strongly localized film. The straight lines are the linear fit to the data between 2 and 20 K. Inset: Logarithm of the anomalous Hall resistance as a function of $T^{-1/3}$ for the same sample.

FIG. 4. (color online). (a) The anomalous Hall resistivity as a function of $\ln T$ for FePt films with 100, 12, 6, 3, 1.6, 1.3, and 1.2 nm thicknesses. (b) The anomalous Hall resistivity as a function of electrical resistivity at 5 K and 2 K. The solid lines are guides to the eye. (c) Log-log plot of the anomalous Hall conductivity as a function of the longitudinal conductivity at 2 K. The solid straight line (red) represents the power law fit to the experimental data of 3-1.2 nm films. The solid curve (blue) sketches the crossover from the intermediate disorder regime to the high disorder regime.

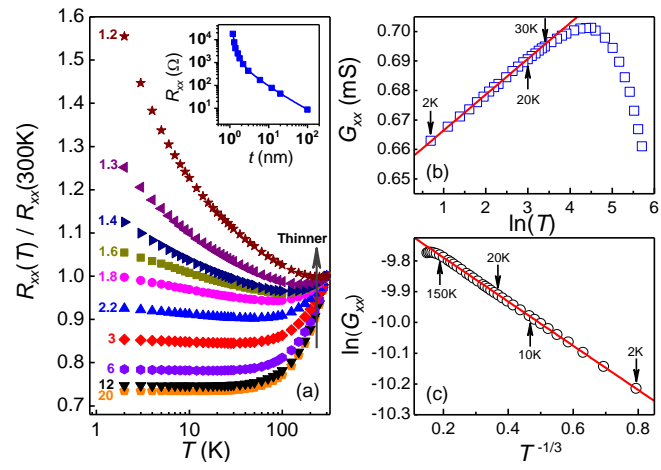

**Figure 1**

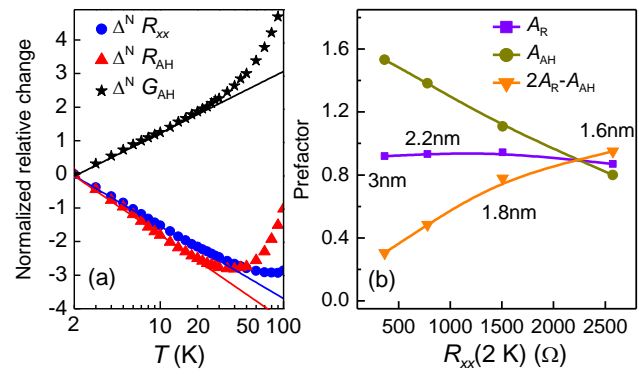

**Figure 2**

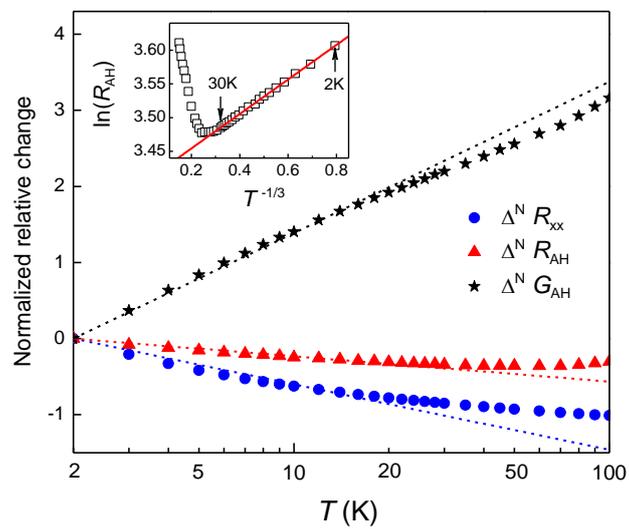

**Figure 3**

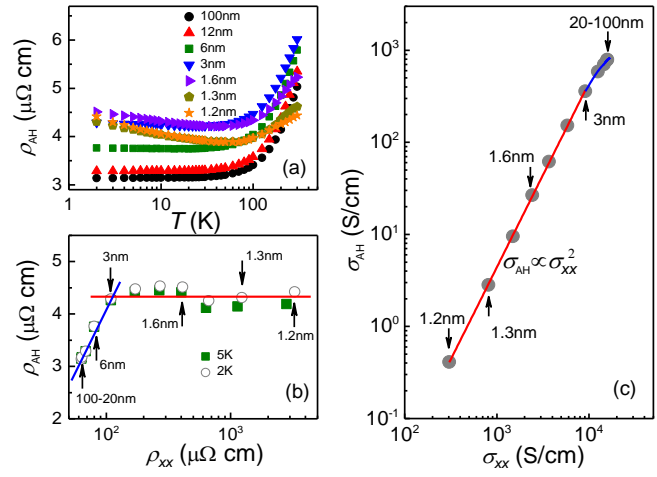